\documentclass[conference,compsoc]{IEEEtran}
\ifCLASSOPTIONcompsoc
  \usepackage[nocompress]{cite}
\else
  \usepackage{cite}
\fi
\ifCLASSINFOpdf
\else
\fi
\usepackage{epsfig,endnotes,multirow,longtable,multicol,tabularx,adjustbox,longtable,pifont,array,subfig,hyperref}
\usepackage{url}
\usepackage{wasysym}
\usepackage{booktabs}
\usepackage{framed}
\usepackage{amssymb}
\newcommand{{\sh}}[0]{smart home}
\newcommand{{\spa}}[0]{smart home personal assistant}
\usepackage[table]{xcolor}
\usepackage{framed}
\colorlet{heatmapbasecolor}{blue!80}
\definecolor{lightgray}{rgb}{0.7, 0.7, 0.7}

\begin{document}
%
\title{\Large \bf SoK: Synthesizing Smart Home Privacy Protection Mechanisms Across Academic Proposals and Commercial Documentations}

\author{
    Under Submission
}


%
\author{\IEEEauthorblockN{Shuning Zhang\IEEEauthorrefmark{1},
Yijing Liu\IEEEauthorrefmark{1},
Yuyu Liu\IEEEauthorrefmark{1}, 
Ying Ma\IEEEauthorrefmark{2},
Shixuan Li\IEEEauthorrefmark{1},\\
Xin Yi\IEEEauthorrefmark{1}\IEEEauthorrefmark{4},
Kanye Ye Wang\IEEEauthorrefmark{3}, 
Qian Wu\IEEEauthorrefmark{1} and
Hewu Li\IEEEauthorrefmark{1}}
\IEEEauthorblockA{\IEEEauthorrefmark{1}Tsinghua University\\
Beijing, China}
\IEEEauthorblockA{\IEEEauthorrefmark{2}University of Melbourne\\ Melbourne, Australia}
\IEEEauthorblockA{\IEEEauthorrefmark{3}University of Macau\\Macau, China}
\IEEEauthorblockA{\IEEEauthorrefmark{4}Corresponding author}}


\maketitle

\begin{abstract}
Pervasive data collection by Smart Home Devices (SHDs) demands robust Privacy Protection Mechanisms (PPMs). The effectiveness of many PPMs, particularly user-facing controls, depends on user awareness and adoption, which are shaped by manufacturers' public documentations. However, the landscape of academic proposals and commercial disclosures remains underexplored. To address this gap, we investigate: (1) What PPMs have academics proposed, and how are these PPMs evaluated? (2) What PPMs do manufacturers document and what factors affect these documentation? To address these questions, we conduct a two-phase study, synthesizing a systematic review of 117 academic papers with an empirical analysis of 86 SHDs' publicly disclosed documentations. Our review of academic literature reveals a strong focus on novel system- and algorithm-based PPMs. However, these proposals neglect deployment barriers (e.g., cost, interoperability), and lack real-world field validation and legal analysis. Concurrently, our analysis of commercial SHDs finds that advanced academic proposals are absent from public discourse. Industry postures are fundamentally reactive, prioritizing compliance via post-hoc data management (e.g., deletion options), rather than the preventative controls favored by academia. The documented protections correspondingly converge on a small set of practical mechanisms, such as physical buttons and localized processing. By synthesizing these findings, we advocate for research to analyze challenges, provide deployable frameworks, real-world field validation, and interoperability solutions to advance practical PPMs.
\end{abstract}


%
\IEEEpeerreviewmaketitle

\section{Introduction}

The proliferation of Internet of Things (IoT) devices in smart homes has introduced substantial privacy risks alongside enhanced functionalities. Users, often not well-versed in the data lifecycle (e.g., collection ~\cite{malkin2018can,crager2017information}, deletion~\cite{cho2020will}), frequently feel powerless, which significantly undermines their trust in these technologies~\cite{schomakers2020understanding}. In response, both academia and industry have developed numerous Privacy Protection Mechanisms (PPMs)--techniques designed to safeguard personal data from unauthorized access and misuse~\cite{sun2021two,jin2022exploring}. These include privacy labels~\cite{Thakkar2022ItWP}, access control systems~\cite{bastys2018if,chi2021pfirewall}, and novel privacy-preserving architectures~\cite{fabiano2017internet}.

\begin{figure}[!htbp]
    \centering
    \includegraphics[width=0.5\textwidth]{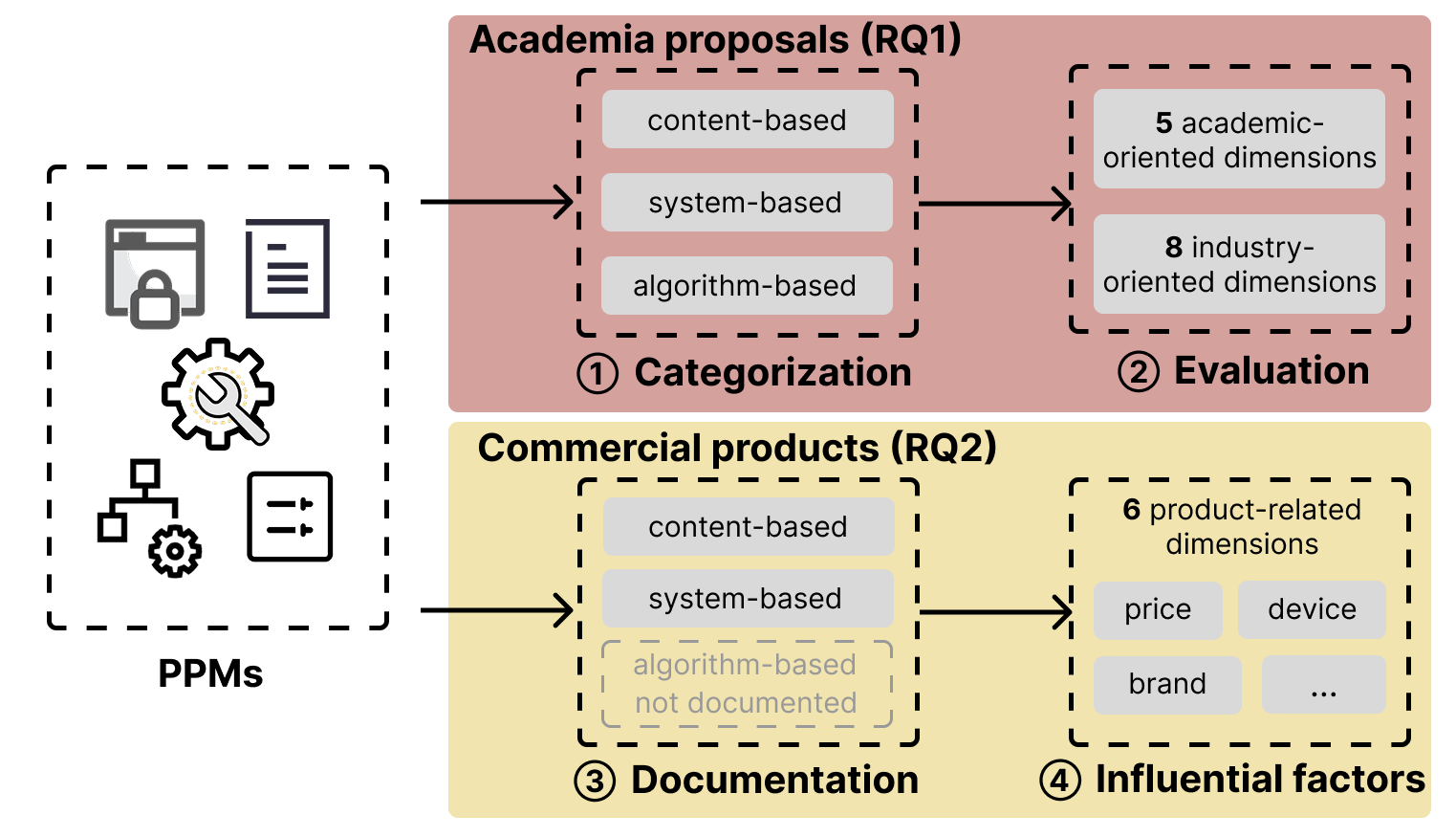}
    \caption{The structure of this paper.}
    \label{fig:framework}
\end{figure}

However, the landscapes of academic PPM proposals and the mechanisms publicly documented by manufacturers for commercial products remain underexplored. Understanding the public-facing commercial landscape is critical, as the efficacy of many PPMs--especially user-facing controls--depends on user awareness and action. Prior research has shown that without clear guidance, users often rely on ad-hoc, physical behaviors, such as unplugging devices, to protect their privacy~\cite{jin2022exploring,lenhart2023you}, hindering the effective use of PPMs. Furthermore, a comprehensive classification of PPMs is lacking, which complicates a rigorous synthesis of academic and commercial approaches. Therefore, to address this gap, this paper systematically reviews and synthesizes PPMs for smart home devices (SHDs) through answering the following research questions (RQs):

$\bullet$ RQ1. What PPMs do the academic papers propose, and how are they evaluated? 

$\bullet$ RQ2. What PPMs do manufacturers publicly disclose in their product-facing documentations, and how do these disclosed PPMs vary across products and manufacturers?

To answer these questions, we conducted a two-phase study, synthesizing 117 top-tier papers and analyzing the public-facing documentation of 86 commercial IoT devices (see Figure~\ref{fig:framework}). Our analysis reveals a disconnect between the two landscapes, not only in proposed solutions but in their validation focus. Regarding RQ1, our review yielded a classification of PPMs into three categories: content-based, system-based, and algorithm-based. Academic research is prolific in proposing novel system- and algorithm-based PPMs. However, these proposals neglect critical deployment barriers such as cost, interoperability, and infrastructure requirements. Furthermore, these technical PPMs lack real-world field validation and analysis of legal compliance.

Towards RQ2, our analysis of the public documentation from 86 commercial devices reveals a different landscape. Beyond the predictable absence of non-user-facing system- or algorithm-based PPMs, we find that even innovative content-based PPMs, such as access control and tangible privacy PPMs, are absent from public discourse. Instead, the publicly documented protections reveal a reactive approach, prioritizing compliance via post-hoc data management (e.g., deletion options), rather than the preventative controls favored by academia. When protections are documented, they converge on a small set of practical mechanisms, including physical affordances (e.g., mute buttons), software-based user control options, and localized processing. Based on these findings, we advocate for research to analyze challenges, provide deployable frameworks, conduct real-world field validation, and offer interoperability solutions to advance practical PPMs. To sum up, this paper has three contributions:

$\bullet$ We provide the first categorization of SHD PPMs, synthesizing 117 academic papers into a content-, system- and algorithm-based taxonomy.

$\bullet$ We conduct a systematic analysis of the academic PPM landscape for SHDs, providing evidence of a focus on technical novelty and a lack of deployment validations.

$\bullet$ We present the first empirical analysis of SHDs' documented privacy protections, identifying the misalignment between documented practices and academic proposals.

\section{Background \& Related Work}

\subsection{Privacy Concerns of IoT Devices}

Smart home technologies offer enhanced convenience but also pose risks related to data misuse \cite{lenhart2023you}. Numerous studies have explored user privacy perceptions regarding smart home technologies \cite{apthorpe2017smart, apthorpe2018discovering,jin2021lean,li2020privacy,malkin2018can,rutledge2016privacy,worthy2016trust,zeng2017end,zheng2018user,zimmermann2018home}. Worthy et al. associated the level of trust in data-collecting entities with users' desires for control over their information, suggesting a direct correlation between lower trust and a higher demand for control \cite{worthy2016trust}. Malkin et al. highlighted uncertainties about how smart TVs manage personal data, what is collected, used, repurposed, and shared \cite{malkin2018can}. Research also indicates that privacy concerns vary with the context of data collection, including consent processes, brands, and types of data \cite{apthorpe2018discovering}. As users are central to SHD usage, some literature proposed user-centric solutions that ensure data control and trust \cite{edu2020smart,zheng2018user}. Regarding concerns and mental models of protection, Zheng et al. noted that some individuals believe their data is meticulously protected by the collectors \cite{zheng2018user}. Collectively, these works highlight the user-centric risks, motivating our systematization of PPMs.

\subsection{PPMs of IoT Devices}

Research in human-computer interaction (HCI) has developed numerous privacy-preserving mechanisms and solutions to enhance user trust and ensure responsible data use in smart homes \cite{albayaydh2022exploring,bahirat2021overlooking,chalhoub2020ux,chalhoub2020innovation,das2018personalized,morgan2022reducing,prange2021priview,wang2023modeling,zheng2018user}. These initiatives aim to create privacy-friendly technologies and policies, while also considering user perceptions on data handling, informing interface designs that enhance individual control over personal data \cite{ahuja2020direction,chalhoub2020factoring,chalhoub2021did,schomakers2020understanding}. Research has also suggested manual restrictions on device functions \cite{malkin2022runtime} and adaptive privacy mechanisms for multi-user settings \cite{zeng2019understanding}. Personalized privacy notifications via IoT devices cater to individual preferences \cite{yao2019defending}, and privacy settings are adjusted based on user scenarios like being home or away \cite{prange2020wish}. Additionally, VR and AR technologies have been proposed for visualizing and safeguarding privacy data in smart homes \cite{schenkluhn2023augmented,george2019investigating}. Despite advancements, most PPMs were proposed in an ad-hoc manner \cite{do2023powering,prange2021priview}, lacking synthesis and evaluation \cite{Thakkar2022ItWP}. Therefore, our work aimed to provide guidance and recommendations for the design of future PPMs.

\subsection{Evaluation and Comparison of PPMs for Smart Homes}

Prior work synthesized and evaluated smart home PPMs from academic and product perspectives. Early research, such as Park et al.~\cite{park2013comparative} examined specific methods like periodic and probabilistic transmission, identifying their trade-offs between anonymity, latency, and energy efficiency. Other syntheses focused on narrow technical categories. For example, Chen et al.~\cite{chen2021survey} summarized data-flow-related PPMs like traffic shaping. Seliem et al.~\cite{seliem2018towards} proposed a broad classification based on four technical classes (e.g., authentication, edge computing) across three architectural levels (device, platform, application). However, this work largely overlooked content-based solutions and neglected user perspectives. More user-centric work, such as Thakkar et al.~\cite{Thakkar2022ItWP}, analyzed privacy notice mechanisms but was similarly limited to a single subset of content-based PPMs. 

A separate body of research has investigated the security and privacy of commercial smart home products, rather than academic proposals. This work often analyzes user-facing artifacts. For example, researchers examined Amazon customer reviews to understand consumer S\&P concerns~\cite{vetrivel2023examining,protick2024unveiling}. Manandhar et al.~\cite{manandhar2022smart} conducted a large-scale analysis of vendor privacy policies, finding them imprecise, difficult to assess, and lacking complete device coverage. Other works conduct in-depth technical security analyses of commercial platforms. These studies have uncovered significant design flaws and vulnerabilities in frameworks like Samsung SmartThings~\cite{fernandes2016security}, platform-wide device interactions~\cite{zhou2019discovering}, specific protocols like JoyLink~\cite{liu2017smart}, and general off-the-shelf device ecosystems~\cite{geneiatakis2017security}. While this body of research provides insights into security vulnerabilities and policies, it does not analyze the documented privacy protections, nor does it contrast these documented protections with academic proposals.

The work most related is from Jin et al.~\cite{jin2022exploring}, who summarized 11 PPMs as storyboards from the perspective of smart home users and subsequently investigated users' opinions of them. Nevertheless, their study focused primarily on mechanisms requiring proactive user engagement. Their classification did not encompass the full range of PPMs, nor did it assess this broader landscape. 

\section{Methodology}\label{three}

To assess the landscape of PPMs, we conducted a two-phase investigation. First, we performed a systematic literature review (SLR) to map the scholarly contributions in the domain. The SLR methodology was chosen for its rigor in consolidating existing knowledge~\cite{mulrow1994systematic} and establishing analytical frameworks~\cite{pare2015synthesizing}. Second, we conducted an empirical analysis of the publicly documented protections of commercially available SHDs. This dual approach allows for a systematic comparison between proposed academic proposals and the publicly documented protections by manufacturers. 

\subsection{Phase 1: Literature Review}

\subsubsection{Review Process}

Our SLR involved four phases: (1) Identification, gathering an initial set of papers using keyword searches; (2) Filtering, applying eligibility criteria to assess the relevance of each paper; (3) Review, entailing detailed reading and categorization of the papers and (4) Analysis, synthesizing and reporting the statistics.

To ensure collection quality, we selected top-tier, peer-reviewed conferences and journals that focus on PPMs. Specifically, we chose the top-10 conferences in privacy and security from the ``Computer Security \& Cryptography'' and ``Human-Computer Interaction'' categories listed in the Google Scholar academic index\footnote{\url{https://scholar.google.com/citations?view_op=top_venues}}. These included notable conferences such as ACM CCS\footnote{ACM Conference on Computer and Communications Security}, IEEE S\&P\footnote{IEEE Symposium on Security and Privacy}, and USENIX Security\footnote{USENIX Security Symposium}. We also included conferences with a strong emphasis on usable privacy and security, or those publishing solid S\&P works, such as SOUPS\footnote{Symposium on Usable Privacy and Security} PETs\footnote{Privacy Enhancing Technologies Symposium}, and ACSAC\footnote{Annual Computer Security Applications Conference}, as well as top-tier IoT venues such as the IEEE Internet of Things Journal, and IoTDI\footnote{The ACM/IEEE International Conference on Internet of Things Design and Implementation}. Our scope covered papers published between 2005 and 2025, ensuring comprehensive coverage of works related to IoT devices. Among these venues, we performed a keyword search in the ACM Digital Library, the IEEE Computer Society Digital Library and Google Scholar, targeting PPM papers related to IoT devices. The search terms used (case-insensitive) were: \textit{(``privacy protection mechanism'' OR ``privacy protection'') AND (``iot'' OR ``internet of things'' OR ``smarthome'' OR ``smart home'')}, with the keywords appearing at least once in the title or abstract. We experimented with adding additional privacy-related terms (e.g., encryption, anonymity, authentication) to our search query. However, this approach yielded very few additional papers, most of which were not relevant to PPMs, so we opted not to include them. In the collection phase, we collected 295 papers in total, with 38 Computer \& Security papers, 31 CHI\footnote{The ACM Conference on Human Factors in Computing Systems} papers, 30 CCS papers, 54 USENIX Security papers and 142 papers from other conferences or journals. 

In the filtering phase, we screened the papers for relevance, excluding those not focused on smart home IoT privacy protection. Exclusion criteria were twofold: (1) papers that did not introduce or evaluate PPMs, including those only introducing attacks, mitigations, or general perceptions of users or other stakeholders, and (2) papers that did not involve smart home IoT devices, such as those exclusively addressing industrial IoT architecture or medical devices. In total, we removed 178 papers, leaving 117. The final set contained 95 full papers, 3 posters and 15 workshop papers. We further analyzed the collected PPMs, as presented in the next section.

\subsubsection{Data Analysis}

To facilitate a comparative analysis with commercial disclosures, we analyzed each paper in two stages: (1) categorizing the proposed PPMs, and (2) coding each paper against a nuanced set of evaluation dimensions.

First, utilizing the topic modeling approach \cite{cooper1988organizing}, we categorized PPMs based on their functionality. We opted against extensive computational analysis, as the scope was limited to well-defined PPMs. We focused on the usage context and existing literature when naming and categorizing PPMs. In the absence of a clear consensus in prior work, names and descriptions were crafted from a technical perspective, ensuring fidelity to the original technology's intent. Methodologically, the first two authors independently analyzed 20\% of the papers, resolving any discrepancies through discussion. They subsequently coded the remaining papers separately, achieving an inter-rater reliability (IRR) score of 0.92. 

Second, each paper was coded against a set of dimensions. Here, a uniform criterion was deemed inappropriate, as different PPMs address distinct problems and are subject to varying constraints. We therefore defined a comprehensive set of academic- and industry-oriented dimensions. We then established a mapping, shown in Table~\ref{tab:ppm_priority_balanced} that identifies the most relevant dimensions for each PPM. For example, \textit{negative UX} is designated as a primary evaluation criterion for content-based PPMs, while \textit{resource constraints} and \textit{infrastructure requirements} are more important for system- and algorithm-based solutions. The full set of dimensions is as follows:

\textbf{For academic-oriented dimensions}, we analyzed the following dimensions to capture aspects central to academic research contributions:

$\bullet$ \textit{A1: Threat model.} The clarity and formality of the specified threat model and attacker capabilities, coded as None (absent), Partial (informally mentioned), or Full (formally defined). Formal threat models are critical as attacks are often highly targeted~\cite{edu2020smart}.

$\bullet$ \textit{A2: Empirical evaluation.} The evaluation methods (e.g., simulation, test-bed, field study, formal proof, user study). This aligns with taxonomies in prior surveys~\cite{stephenson2022sok}.

$\bullet$ \textit{A3: Data lifecycle coverage.} The data flow stages (i.e., collection, transmission, processing, storage, deletion) covered by the PPM (multi-label binary). This follows research highlighting the importance of analyzing data practices at each stage of the lifecycle~\cite{solove2005taxonomy,ma2025privacy}.

$\bullet$ \textit{A4: Privacy property targeted.} The core privacy properties the PPM aims to enhance (e.g,. confidentiality, anonymity, control), using the established terminology from Pfitzmann~\cite{pfitzmann2010terminology}.

$\bullet$ \textit{A5: Resource constraints.} Consideration of resource constraints (e.g., CPU, memory, energy), coded as None (not stated), Partial (mentioned), or Full (formally quantified). Modeling these is important for deployment~\cite{lin2016iot}.

\textbf{For industry-oriented dimensions,} to understand the extent to which academic research considers factors critical for industry adoption, the papers were also coded using dimensions reflecting typical industry concerns:

$\bullet$ \textit{I1: Cost.} Consideration of financial or resource cost, coded as None (not discussed), Partial (mentioned), or Full (quantitatively analyzed). This addresses practical deployment feasibility especially in companies~\cite{acar2016you}.

$\bullet$ \textit{I2: Legal compliance.} Explicit references to legal standards (e.g., GDPR, CCPA, PIPL). This addresses the critical regulatory compliance issues emphasized by prior work~\cite{lee2024don,yang2025guidelines,horstmann2025sorry}.

$\bullet$ \textit{I3: Negative user experience (UX).} Evaluation of negative UX impacts (e.g., latency, friction), coded as None, Partial (discussed), or Full (measured). This focus is critical, as usability problems and frictions are known barriers~\cite{jin2022exploring}.

$\bullet$ \textit{I4: Incentive/positive UX.} Consideration of positive user incentives (e.g., improved experience, purchase intention), coded as None, Partial (qualitative discussion) or Full (quantitative). This focus is motivated by research indicating that subjective experience with privacy influences users' willingness of adoption~\cite{borgert2025value}.

$\bullet$ \textit{I5: Integration depth.} The PPM's operational layer (e.g., firmware, app, cloud, mixed). This design choice is vital as it often dictates PPMs' effectiveness and constraints~\cite{vetrivel2023examining,lin2016iot}.  

$\bullet$ \textit{I6: Infrastructure requirement.} Requirement for additional hardware (e.g., gateways, edge servers), coded as a binary attribute. This architectural analysis is consistent with prior SoKs~\cite{zavalyshyn2022sok}.

$\bullet$ \textit{I7: Interoperability conflicts.} Discussed conflicts with existing standards (e.g., Zigbee, Matter), coded as a binary attribute. This addresses a known research challenge~\cite{singh2023intercompatibility}.

$\bullet$ \textit{I8: Scalability evidence.} Demonstration of scalability (e.g., regarding device numbers), coded as None, Partial (mentioned), or Full (measured), which is a common deployability goal~\cite{reichherzer2016case,javed2020scalable}

The coding was performed by 2 researchers, where they first aligned their understanding of the codebook and coding criteria. They then selected 20 papers as a subset, coded these papers and calculated the IRR. The IRR, calculated using Krippendorff's alpha, reached 0.85. They resolved disagreements through discussion. They then independently coded half of the remaining papers\footnote{See Appendix~\ref{app:dimension} and \href{https://github.com/anonymous-sp-sub/anonymous-sp-submission}{anonymous repo} for the detailed definitions for each coding dimension and coding criteria.}.

\begin{table}[h!]
\centering
\caption{Prioritized evaluation dimensions for each PPM categories (highlighted with checkmarks).}
\label{tab:ppm_priority_balanced}
\resizebox{\columnwidth}{!}{%
\begin{tabular}{@{}lccc@{}} 
\toprule
\textbf{Dimension} & \textbf{Content-based} & \textbf{System-based} & \textbf{Algorithm-based} \\
\midrule
\multicolumn{4}{l}{\textit{Academic-Oriented Dimensions}} \\
A1: Threat Model & & \checkmark & \checkmark \\
A2: Empirical Evaluation & \checkmark & \checkmark & \checkmark \\
A3: Data Lifecycle & \checkmark & \checkmark & \checkmark \\
A4: Privacy Property & \checkmark & \checkmark & \checkmark \\
A5: Resource Constraints & & \checkmark & \checkmark \\
\midrule
\multicolumn{4}{l}{\textit{Industry-Oriented Dimensions}} \\
I1: Cost & & \checkmark & \checkmark \\
I2: Legal Compliance & \checkmark & \checkmark & \checkmark \\
I3: Negative UX & \checkmark & & \\
I4: Positive UX & \checkmark & & \\
I5: Integration Depth & & \checkmark & \checkmark \\
I6: Infrastructure & & \checkmark & \checkmark \\
I7: Interoperability & & \checkmark & \\
I8: Scalability & & \checkmark & \checkmark \\
\bottomrule
\end{tabular}
}%
\end{table}

\subsection{Phase 2: Analysis of Commercial Privacy Disclosures}\label{six}

To bridge academic innovation with market reality, our second phase analyzes the publicly documented privacy protections of 86 commercial smart home devices. We deliberately focus on the public disclosure--the stated features, policies, and design choices (e.g., in product descriptions, privacy portals, and whitepapers)--as this constitutes the primary, and often only, information available to consumers and regulators. This public narrative directly shapes user risk perception, guides purchase decisions, and influences the market incentives for privacy~\cite{hunte2024effect}.

\subsubsection{Collection Process}

We collected implementation details using the search query (case non-sensitive) \textit{(``internet of things'' OR ``iot'') AND (``smart home'' OR ``smarthome'')}\footnote{We tried adding other keywords like `intelligent home', `smart space' but did not acquire new results.}, retrieving web pages from the top 100 search results in September 2025. Data collection ceased when no new products were identified from additional websites. Two primary authors collaboratively screened and extracted details on the brand, model, and descriptions (including names) of IoT devices, filtering out those not intended for smart home use (e.g., industrial devices). In this process, they frequently discussed to resolve any disagreements.

Subsequently, they searched for these device names on official websites to identify the PPMs documented for each device, ensuring consistency with the methods users might employ to obtain PPM information. They also consulted the corresponding documentation (e.g., whitepaper\footnote{for example, Xiaomi's whitepaper: \url{https://trust.mi.com/docs/iot-privacy-white-paper-global/}}) from the companies if retrievable. If a specific PPM was implemented but not described on the official website or in the statements from the companies, it was excluded from our analysis. Our analysis reflects not the ground-truth implementation but its public disclosure. This approach aligns with the user's perspective, as information not explicitly presented would reasonably be perceived as absent by a consumer making a purchase decision. 

We adopted qualitative analysis due to the unstructured nature of PPMs. The same two primary authors jointly coded the websites for 10 IoT devices and set the coding criteria, with intermittent discussion to resolve disagreements. They then separately coded the rest of the websites. The IRR using Cohen's kappa reached 0.95. Because the codebook, coding process could be reflected in the presented results, we chose not to include them separately.

We examined a total of 86 types of IoT devices. The specific devices were aggregated at two different levels: the manufacturers such as Google, Amazon and their specific models such as Google Camera, and Google Hub. Notably, we followed previous literature and product classification in classifying the models~\cite{kumar2019all}. This resulted in 46 manufacturers and 61 models (note that one model could have multiple types, like Nest Cam indoor, Nest Cam with floodlight, Nest Cam battery, etc.), as summarized in Table~\ref{tbl:implementation_situation}. 

\subsubsection{Data Analysis}

With the same qualitative analysis approach, we coded the PPMs each product documented. Instead of simply scoring the documentation, two authors coded the specific, nuanced description of how each PPM was documented. For example, rather than coding \textit{Access Control} as ``documented'', we coded the description (e.g., \textit{``provides an in-app button for data deletion''} or \textit{``allows data review after 72 hours''}). This approach preserves the details of real-world implementations. The two authors jointly coded an initial 10 devices to establish the codebook, independently coded the remaining devices, and resolved all disagreements through discussion to reach consensus. Furthermore, to understand the drivers of these documented privacy protections, we coded each product across a set of dimensions informed by prior work~\cite{emami2023consumers,kumar2019all,vetrivel2023examining}, organized into two thematic groups: product and market context (including \textit{price}, \textit{primary market} and \textit{brand's country of origin}), and academic-oriented dimensions (including \textit{A1: Threat Model}, \textit{A3: Data Lifecycle Coverage}, and \textit{A4: Privacy Property Targeted}), which are adapted from criteria used for evaluating academic papers (see Appendix~\ref{app:product_dimension} for the justification of these dimensions). 

\subsection{Limitations}

Our methodology has several limitations. Our systematic literature review was confined to a defined set of top-tier, English-language academic venues, potentially omitting relevant PPMs from workshops, alternative venues, or non-English publications. Similarly, our commercial sample, derived from the top 100 Google search results, inherently favors popular, well-marketed devices, potentially under-representing niche or non-English-market products. Furthermore, our analysis of commercial privacy protection documentation is a temporal snapshot from September 2025 and may not capture subsequent documentation updates in this rapidly evolving market. Finally, this study was constrained by a lack of physical access to the devices, which prevented the retrieval of PPMs documented on physical materials, such as user manuals or on-box ``privacy nutrition labels''. Future work should aim for a more comprehensive analysis to address these constraints.

\section{Analysis of Academic Research}

Based on 117 academic papers, we analyze key dimensions of proposed PPMs, such as their classification, threat models, empirical evaluations, legal compliance and integration depths. The analysis reveals a substantial growth in PPM research, with 80.3\% (94/117) of surveyed literature published between 2018 and 2023. The years 2022 and 2023 (25 papers each year) were particularly prolific, underscoring an increasing focus on SHD privacy challenges. 

This academic innovation is driven almost exclusively by universities, which constituted over 90\% of publications in most years, while industry collaborations remain rare (e.g., 7.7\% in 2021 and 4.0\% in 2022). 81/117 papers focus on the design and proposal of novel PPMs, such as Emami-Naeini et al's exploration of the design space for IoT privacy labels. 11/117 papers derived proposals from qualitative analyses of user privacy concerns. 22/117 papers focused on evaluating or surveying existing PPMs. However, these evaluative studies neither compared academic proposals with industry documentation nor synthesized these landscapes.

\begin{table*}[t!]
\centering
\caption{Analysis of PPMs in academic literature. Values represent frequency counts unless otherwise specified. DN, TP, PL, AC, SD, AU, BL, DP, TR denoted dynamic notification, tangible privacy, privacy label, access control, system design, authorization, blockchain, differential privacy and trading respectively. N/P/F refers to Not Stated(None)/Mentioned(Partial)/Formally Quantified(Full). Prioritized dimensions are shown in black, while others are in light gray. Note that one paper may have multiple PPMs, evaluation types, data lifecycle coverage, and targeted properties.}
\label{tab:ppm_summary}
\resizebox{\textwidth}{!}{%
\begin{tabular}{@{} l l *{9}{c} @{}}
\toprule
& & \multicolumn{4}{c}{\textbf{Content-based}} & \multicolumn{2}{c}{\textbf{System-based}} & \multicolumn{3}{c}{\textbf{Algorithm-based}} \\
\cmidrule(lr){3-6} \cmidrule(lr){7-8} \cmidrule(lr){9-11}
\textbf{Dimension} & \textbf{Value} & \textbf{DN} & \textbf{TP} & \textbf{PL} & \textbf{AC} & \textbf{SD} & \textbf{AU} & \textbf{BL} & \textbf{DP} & \textbf{TR} \\
\midrule


\multicolumn{2}{@{}l}{Num of Papers} & 16 (13.7\%) & 11 (9.4\%) & 11 (9.4\%)& 29 (24.8\%) & 68 (58.1\%) & 26 (22.2\%) & 22 (18.8\%) & 10 (8.5\%) & 4 (3.4\%)\\
\midrule

\multicolumn{11}{@{}l}{\textbf{Academic-Oriented Dimensions}} \\
\textbf{A1:} Threat Model &  & \textcolor{lightgray}{7 (43.8\%)} & \textcolor{lightgray}{5 (45.5\%)} & \textcolor{lightgray}{4 (36.4\%)} & \textcolor{lightgray}{21 (72.4\%)} & 57 (83.8\%) & 24 (92.3\%) & 20 (90.9\%) & 10 (100.0\%) & 2 (50.0\%) \\
\midrule
\multirow{5}{*}{\textbf{A2:} {Empirical Evaluation}}
 & Simulation & 0 & 0 & 0 & 6 & 18 & 9 & 10 & 5 & 0 \\
 & Test-bed & 0 & 4 & 0 & 9 & 23 & 8 & 6 & 3 & 1 \\
 & Field Study & 0 & 0 & 0 & 2 & 1 & 0 & 1 & 0 & 1 \\
 & Formal Proof & 0 & 0 & 0 & 0 & 3 & 1 & 2 & 0 & 0 \\
 & User Study & 9 & 7 & 7 & 5 & 8 & 1 & 0 & 0 & 0 \\
\midrule
\multirow{5}{*}{\textbf{A3:} \parbox{2.5cm}{Data Lifecycle Coverage}}
 & Collection &     15 & 11 & 10 & 26 & 60 & 21 & 19 & 7  & 3 \\
 & Transmission &   15 & 9  & 10 & 28 & 60 & 26 & 20 & 10 & 4 \\
 & Storage &        11 & 5  & 7  & 22 & 48 & 19 & 17 & 5  & 3 \\
 & Processing &     14 & 7  & 10 & 27 & 59 & 24 & 20 & 9  & 3 \\
 & Deletion &       11 & 2  & 7  & 8  & 20 & 10 & 6  & 0  & 1 \\
\midrule
\multirow{8}{*}{\textbf{A4:} \parbox{2.5cm}{Privacy Property Targeted}}
 & Confidentiality      & 11 & 8  & 6  & 26 & 57 & 26 & 20 & 9 & 3 \\
 & Unlinkability        & 2  & 3  & 1  & 8  & 28 & 14 & 10 & 6 & 2 \\
 & Minimisation         & 11 & 6  & 7  & 20 & 40 & 7  & 10 & 3 & 2 \\
 & Transparency         & 16 & 10 & 11 & 21 & 34 & 8  & 11 & 0 & 2 \\
 & Control              & 16 & 10 & 11 & 26 & 49 & 13 & 16 & 2 & 4 \\
 & Accountability       & 13 & 5  & 8  & 13 & 24 & 9  & 11 & 1 & 2 \\
 & Differential Privacy & 1  & 0  & 1  & 3  & 11 & 0  & 6  & 5 & 0 \\
 & Anonymity            & 4  & 2  & 2  & 8  & 28 & 11 & 11 & 7 & 3 \\
\midrule
\textbf{A5:} Resource Constraints & (N/P/F) & \textcolor{lightgray}{15/1/0} & \textcolor{lightgray}{7/4/0} & \textcolor{lightgray}{9/2/0} & \textcolor{lightgray}{14/14/1} & 23/41/4 & 5/21/0 & 7/14/1 & 1/8/1 & 3/1/0 \\
\midrule
\multicolumn{11}{@{}l}{\textbf{Industry-Oriented Dimensions}} \\
\textbf{I1:} Cost & (N/P/F) & \textcolor{lightgray}{12/4/0} & \textcolor{lightgray}{7/2/2} & \textcolor{lightgray}{6/4/1} & \textcolor{lightgray}{19/3/7} & 40/13/15 & 11/5/10 & 7/7/8 & 2/4/4 & 2/0/2 \\
\midrule
\multirow{5}{*}{\textbf{I2:} \parbox{2.5cm}{Legal Compliance}}
 & GDPR  & 10 & 1 & 7 & 6 & 10 & 1 & 4 & 0 & 1 \\
 & CCPA  & 2  & 1 & 2 & 0 & 2  & 0 & 0 & 0 & 0 \\
 & COPPA & 4  & 0 & 2 & 1 & 3  & 0 & 1 & 0 & 1 \\
 & HIPAA & 2  & 0 & 2 & 1 & 5  & 0 & 0 & 0 & 0 \\
 & PIPL  & 3  & 1 & 2 & 0 & 3  & 0 & 0 & 0 & 0 \\
\midrule
\textbf{I3:} Negative UX & (N/P/F) & 10/4/2 & 6/4/1 & 7/3/1 & 15/6/8 & \textcolor{lightgray}{31/17/20} & \textcolor{lightgray}{11/5/10} & \textcolor{lightgray}{7/5/10} & \textcolor{lightgray}{3/2/5} & \textcolor{lightgray}{3/1/0} \\
\midrule
\textbf{I4:} Incentive/Positive UX & (N/P/F) & 7/7/2 & 6/4/1 & 4/6/1 & 11/14/4 & \textcolor{lightgray}{35/29/4} & \textcolor{lightgray}{13/7/6} & \textcolor{lightgray}{7/9/6} & \textcolor{lightgray}{4/4/2} & \textcolor{lightgray}{1/2/1} \\
\midrule
\multirow{6}{*}{\textbf{I5:} \parbox{2.5cm}{Integration Depth}}
 & Mixed          & \textcolor{lightgray}{6} & \textcolor{lightgray}{1} & \textcolor{lightgray}{4} & \textcolor{lightgray}{9} & 24 & 12 & 12 & 3 & 1 \\
 & Companion App  & \textcolor{lightgray}{6} & \textcolor{lightgray}{3} & \textcolor{lightgray}{4} & \textcolor{lightgray}{8} & 10 & 2  & 3  & 2 & 2 \\
 & Cloud          & \textcolor{lightgray}{2} & \textcolor{lightgray}{1} & \textcolor{lightgray}{1} & \textcolor{lightgray}{6} & 15 & 10 & 6  & 3 & 1 \\
 & Firmware       & \textcolor{lightgray}{0} & \textcolor{lightgray}{2} & \textcolor{lightgray}{0} & \textcolor{lightgray}{1} & 4  & 0  & 0  & 0 & 0 \\
 & Standalone App & \textcolor{lightgray}{0} & \textcolor{lightgray}{0} & \textcolor{lightgray}{0} & \textcolor{lightgray}{1} & 0  & 0  & 0  & 0 & 0 \\
 & Embedded OS    & \textcolor{lightgray}{1} & \textcolor{lightgray}{2} & \textcolor{lightgray}{1} & \textcolor{lightgray}{1} & 13 & 1  & 1  & 1 & 0 \\
\midrule
\multirow{7}{*}{\textbf{I6:} \parbox{2.5cm}{Infrastructure Requirement}}
 & Edge Server        & \textcolor{lightgray}{1} & \textcolor{lightgray}{1} & \textcolor{lightgray}{1} & \textcolor{lightgray}{5} & 17 & 9 & 5 & 4 & 2 \\
 & Hub                & \textcolor{lightgray}{3} & \textcolor{lightgray}{1} & \textcolor{lightgray}{1} & \textcolor{lightgray}{4} & 7  & 1 & 1 & 0 & 0 \\
 & Extra Gateway      & \textcolor{lightgray}{1} & \textcolor{lightgray}{2} & \textcolor{lightgray}{0} & \textcolor{lightgray}{2} & 11 & 7 & 1 & 1 & 0 \\
 & Coordinator        & \textcolor{lightgray}{1} & \textcolor{lightgray}{2} & \textcolor{lightgray}{1} & \textcolor{lightgray}{3} & 6  & 2 & 2 & 2 & 0 \\
 & Federated Learning & \textcolor{lightgray}{0} & \textcolor{lightgray}{1} & \textcolor{lightgray}{0} & \textcolor{lightgray}{1} & 4  & 0 & 4 & 1 & 0 \\
 & Crypto-support     & \textcolor{lightgray}{0} & \textcolor{lightgray}{0} & \textcolor{lightgray}{0} & \textcolor{lightgray}{1} & 3  & 1 & 1 & 0 & 0 \\
 & Others             & \textcolor{lightgray}{1} & \textcolor{lightgray}{3} & \textcolor{lightgray}{0} & \textcolor{lightgray}{3} & 5  & 7 & 1 & 1 & 1 \\
\midrule
\textbf{I7:} \parbox{2.5cm}{Interoperability \\Conflicts} & & \textcolor{lightgray}{1} & \textcolor{lightgray}{2} & \textcolor{lightgray}{0} & \textcolor{lightgray}{1} & 7 & 0 & \textcolor{lightgray}{1} & \textcolor{lightgray}{1} & \textcolor{lightgray}{0} \\
\midrule
\textbf{I8:} Scalability Evidence & (N/P/F) & \textcolor{lightgray}{15/1/0} & \textcolor{lightgray}{11/0/0} & \textcolor{lightgray}{10/1/0} & \textcolor{lightgray}{24/5/0} & 60/6/2 & 24/2/0 & 17/4/1 & 7/2/1 & 4/0/0 \\
\bottomrule
\end{tabular}%
} 
\end{table*}

\subsection{Categorization of PPMs}\label{sec:categorization}

We categorized the PPMs from 117 papers into 9 types, which are grouped into 3 overarching categories. Unlike prior work that classifies by architectural layer~\cite{seliem2018towards}, our taxonomy is based on the solution's core approach (e.g., system-level or algorithm-level). This framework, similar to user-, data- and network-centric classifications~\cite{winkler2014security}, accommodates emerging techniques like blockchain-based PPMs~\cite{winkler2014security}.

$\bullet$ \textbf{Content-based PPMs (56/117 papers).}

(1) \textit{Dynamic Notification (16/117 papers)}: Methods
that provide real-time alerts about data collection or privacy-related activities, often manifesting as visual dashboards~\cite{windl2022saferhome,thakkar2022would} or notifications~\cite{mhaidli2020listen} to keep users informed. 

(2) \textit{Tangible Privacy (11/117 papers)}: Physical, typically hardware-based, mechanisms for direct user control over sensor activity. This includes, for example, microphones that are only powered when manually activated by the user~\cite{do2023powering}.

(3) \textit{Privacy Label (11/117 papers)}: Visual indicators designed to inform users about a device's data collection and processing practices. These range from standardized privacy ``nutrition'' labels~\cite{emami2020ask} to privacy notices~\cite{schaub2015design,colnago2020informing}. 

(4) \textit{Access Control (29/117 papers)}: Tools that allow users to manage and control personal data flow. These are often implemented as management systems for customizing data collection settings, frequently using rule-based frameworks~\cite{bastys2018if,he2018rethinking}, such as IFTTT-like rules~\cite{xu2019privacy}. 

$\bullet$ \textbf{System-based PPMs (84/117 papers).}

(5) \textit{System Design (68/117 papers)}: Proposals for novel privacy-preserving system architectures. These solutions are often tailored for IoT constraints and differ from other defined categories (e.g., authorization, differential privacy, blockchains), such as by adopting fog computing or optimizing on-device processing for privacy~\cite{arruda2019toward,gupta2021phin}.

(6) \textit{Authorization (26/117 papers)}: Novel systems for authentication and access control to isolate and manage sensor data. These methods focus on ensuring only authorized users can access sensitive information, thereby minimizing unauthorized data collection~\cite{wilson2018digital,ali2019internet}.

$\bullet$ \textbf{Algorithm-based PPMs (32/117 papers).}

(7) \textit{Blockchain (22/117 papers)}: The use of decentralized blockchain infrastructure to protect user privacy. These systems aim to enable secure, anonymized data sharing by removing centralized data holders~\cite{awan2019poster,qiu2020novel}.

(8) \textit{Differential Privacy (10/117 papers)}: Techniques applying k-anonymity, differential privacy or similar algorithms to protect privacy. These methods use data obfuscation to allow users to access services without disclosing identifiable personal information~\cite{yin2017location}.

(9) \textit{Trading (4/117 papers)}: Market-based methods that involve pricing and exchanging private information between users and data buyers. This approach uses pricing and bidding mechanisms to facilitate protected data sharing, operating on the assumption that users will willingly share data if the transaction is fair.

The names for these types, except \textit{System Design}, were drawn from existing literature. We introduce \textit{System Design} as a category to encompass works focusing on novel SHD architectures. The three high-level categories (content-, system-, and algorithm-based) classify PPMs by their approach to privacy protection. We avoid creating composite classes to facilitate direct comparison across PPMs.

\subsection{Analysis From Academic-Oriented Perspective}

\subsubsection{Threat Modeling and Rigor}

\textbf{Our analysis reveals a disparity in methodological rigor, with content-based PPMs frequently proposed without the formal threat models common in system-based PPMs.} While a majority of the surveyed literature (31.6\%, 37 papers) provides an explicit threat model and 44.4\% (52 papers) rely on implicit assumptions, this rigor is heavily concentrated and inconsistent. Moreover, this rigor correlates strongly with PPM categories: most system- and algorithm-based PPMs have formal threat models (e.g., 100\% of \textit{Differential Privacy}, 92\% of \textit{Authorization}, and 84\% of \textit{System Design} papers defined a threat model). Conversely, this formality is notably less prevalent in content-based mechanisms (44\% for \textit{Dynamic Notification}, 45\% for \textit{Tangible Privacy}, and 36\% for \textit{Privacy Labels}). This gap suggests that many content-based PPMs are proposed without a fully articulated model of the threats they purport to mitigate.

\textbf{When threat models are specified, their nature also varies from informal models to formal frameworks.} For instance, Tan et al.~\cite{tan2023you} defined an informal threat model, assuming an adversary who exploits traffic-analysis-based IoT fingerprinting to trace user activities. In contrast, Dong et al.~\cite{dong2020your} present a formal approach, proposing and evaluating a complete traffic analysis framework. They operationalized an attack based on sequence-learning techniques (e.g., LSTM) that leveraged temporal packet relations for device identification, rigorously evaluating its high accuracy under different environmental settings. 

\subsubsection{Evaluation Methodologies}

\textbf{Our analysis reveals that evaluation is dominated by technical validation in controlled settings, with the test-bed being the most prevalent method.} The \textit{test-bed} (31.6\%, 37 papers) is often used for large-scale analysis. For example, Moghaddam et al.~\cite{mohajeri2019watching} employed a smart crawler test-bed to analyze over 2,000 OTT channels, demonstrating pervasive tracking on \textit{Roku} and \textit{Amazon Fire TV}. Similarly, Edu et al.~\cite{edu2021skillvet} developed SkillVet to evaluate 199,295 Alexa skills, finding 43\% followed poor privacy practices.

\textbf{Secondary methods like simulation are common for evaluating large-scale algorithm-based PPMs, while formal proofs, as expected, are rare.} \textit{Simulation} (23.1\%, 27 papers) enables scalability evaluation, as when Yu et al.~\cite{yu2024highly} validated their FACT+ access control system with 12.8 million synthetic users across real-world scenarios. \textit{Formal proof} (2.6\%, 3 papers) is rare, utilized primarily for system-based mechanisms. 

\textbf{We observe a clear methodological split across PPM type, and a critical scarcity of real-world field studies.} \textit{User studies} (14.5\%, 17 papers) are the primary domain for \textit{content-based PPMs} like \textit{Dynamic Notification} (5 papers), \textit{Privacy Labels} (7 papers), and \textit{Tangible Privacy} (7 papers). For instance, Ahmad et al.~\cite{ahmad2022tangible} used a between-subjects experiment to find tangible controls were perceived as more trustworthy and usable. In sharp contrast, \textit{system-based PPMs} are validated almost exclusively by technical evaluations, such as \textit{test-beds} and \textit{simulation} (e.g., 39 papers for \textit{System Design} and 17 for \textit{Authorization}). Critically, \textit{field studies} (2.6\%, 3 papers) are markedly rare, highlighting a gap in demonstrated real-world applicability.

\subsubsection{Data Lifecycle of Protection}

\textbf{Our analysis (Table~\ref{tab:ppm_summary}) reveals a heavy concentration on active data stages, while end-of-life stages are largely neglected.} The most frequently addressed stages are \textit{data transmission} (92.3\%, 108 papers) and \textit{processing} (89.7\%, 105 papers), a focus characteristic of system-based approaches like \textit{System Design} (60 papers mentioning transmission and processing respectively). For example, Hou et al.~\cite{hou2021model} concentrate exclusively on the processing phase, proposing a system to secure ML model inference within a trusted enclave on an edge device. \textit{Data collection} is also a primary target (86.3\%, 101 papers), particularly for content-based PPMs like \textit{Dynamic Notification} (30 papers). This stage is sometimes addressed at the hardware level. Wang et al.~\cite{wang2023modeling}, for instance, proposed low-resolution sensors to inherently limit visual information capture.

\textbf{Conversely, \textit{data deletion} is the most neglected stage, constituting an oversight.} Despite a few notable exceptions, such as Habib et al.'s~\cite{habib2022evaluating} evaluation of interfaces for opt-out and deletion, \textit{data deletion} is addressed in only 29.9\% (35 papers). This gap suggests that many academic designs are not explicitly engaging with regulatory principles like the GDPR's ``right to be forgotten''.

\subsubsection{Core Privacy Objectives}

\textbf{Research targets a variety of privacy goals, prioritizing technical \textit{confidentiality} (85.5\%, 100 papers) and user \textit{control} (71.8\%, 84 papers).} The emphasis on \textit{confidentiality} is often realized via system architectures. For instance, Lian et al.~\cite{lian2022decentralized} use federated learning to train models locally, thereby avoiding data sharing. A secondary tier of objectives includes \textit{minimisation} and \textit{transparency} (both 53.8\%, 63 papers). Thakkar et al.~\cite{thakkar2022would}, for example, linked \textit{transparency} to \textit{control}, emphasizing that awareness alone is insufficient, as users ``also expect to have control'' to avoid ``a sense of helplessness''. Lian et al.~\cite{lian2022decentralized} also addressed \textit{minimisation} by exchanging only model parameters. Less frequent objectives include \textit{accountability} (40.2\%, 47 papers) and \textit{anonymity} (38.5\%, 45 papers). For instance, Apthorpe et al.~\cite{apthorpe2019evaluating} evaluated how \textit{transparency} (notification) and \textit{control} (data deletion) align with parental norms while also addressing \textit{confidentiality}, \textit{minimisation} (via storage limits) and \textit{accountability} (complying with COPPA).

\textbf{We observe a division of goals across PPMs where content-based PPMs stress user agency and system-based PPMs highlight technical goals.} User-centric goals like \textit{transparency} are primarily addressed by \textit{content-based} PPMs (\textit{Dynamic Notification}, 16 papers; \textit{Access Control}, 21 papers). In contrast, technical properties like \textit{confidentiality} and \textit{anonymity} are driven by \textit{system-based} architectures (\textit{System Design}, 57 papers for \textit{confidentiality}). This suggests that while \textit{system-based} designs build the technical foundation, \textit{content-based} interfaces focus more on user agency.

\subsubsection{Resource Constraints}

\textbf{Our analysis finds that, while resource constraints are critical for deployable \textit{system-based} and \textit{algorithm-based} PPMs, the literature mostly discusses this dimension, with formal quantification being exceptionally rare.} A majority of papers (59.8\%, 70 papers) mention this issue, while a significant portion (40.2\%, 47 papers) do not. This acknowledgment is concentrated in relevant categories like \textit{System Design} (44 papers), \textit{Authorization} (21 papers), \textit{Blockchain} (15 papers), and \textit{Differential Privacy} (9 papers). However, this analysis is typically qualitative rather than rigorous quantification. For instance, Majeed et al.~\cite{majeed2022rectification} qualitatively analyze constraints by identifying the high computational cost of differential privacy as a key limitation. Formal evaluation of resource impact is exceptionally rare (3.4\%, 4 papers). A notable exception is Naor et al.~\cite{naor2019not}, who implemented their protocol on a Raspberry Pi 3 to measure processing run times and model trade-offs for ``low-power devices''. Conversely, \textit{content-based} PPMs, for which this is a less critical dimension, largely overlook this issue (e.g., unstated in 15/16 \textit{Dynamic Notification} papers). 

\subsection{Analysis From Industry-Oriented Perspective}

\subsubsection{Cost Considerations}

\textbf{We find that cost, a primary constraint for commercial viability, is frequently overlooked in the academic literature. Even considered, the analysis is often qualitative.} A majority of the surveyed literature (55.6\%, 65 papers) did not address implementation or operational cost in any way. When cost is considered, the analysis is often qualitative rather than quantitative. Only 27 papers presented a quantitative analysis, which ranged from computational or communication overhead to direct hardware expenditure. For example, Lu et al.~\cite{lu2020blockchain} frame their privacy-preserving mechanism as a ``low-cost'' solution, quantifying this by its avoidance of the significant ``communication overhead'' associated with traditional centralized exchanges. In contrast, Wang et al.~\cite{wang2022camshield} provide a direct bill-of-materials analysis for their ``bolt-on'' companion device, explicitly stating its prototype cost (\$150) and noting the potential for low-cost alternatives.

\textbf{This omission of cost is more severe in system-based categories where deployment cost is critical, suggesting a disconnect from real-world constraints.} A majority of papers on \textit{System Design} (59\%) and \textit{Access Control} (64\%) fail to discuss any cost implications. \textit{Algorithm-based} PPMs demonstrate a higher degree of consideration, with formal cost quantification present in 50\% of \textit{trading-based} and 40\% of \textit{differential privacy} studies. 

\subsubsection{Legal and Regulatory Compliance}

\textbf{Our analysis finds that explicit engagement with legal frameworks is sparse, heavily skewed toward GDPR, and largely absent from papers proposing system-based PPMs.} A significant majority of papers (70.9\%, 83 papers) made no mention of regulatory compliance. Among the minority that did, the EU's GDPR was the most frequently cited, referenced in 21 papers (17.9\%), particularly in studies on \textit{Dynamic Notification} (10 mentions) and \textit{System Design} (9 mentions). References to other regulations are minimal: COPPA was mentioned in only 6 papers (5.1\%) and CCPA in just 4 (3.4\%). Notably, papers proposing \textit{Differential Privacy} and \textit{Blockchain} made very few explicit references to these legal standards, indicating a potential disconnect between novel technical approaches and regulation.

\textbf{When regulations are cited, the engagement is used to ground design principles rather than for deep legal analysis.} For instance, Feng et al.~\cite{feng2021design} ground their design space for privacy choices in GDPR and CCPA, briefly referencing high-level usability concepts from GDPR Article 7, such as consent being ``freely given'' and ``intelligible''. 

\subsubsection{Impact on User Experience: Friction and Incentives}

\textbf{Our analysis reveals that both the usability friction and the incentive structures of proposed PPMs are inconsistently evaluated.} 55/117 papers (47.0\%) did not mention any potential drawbacks, suggesting that usability costs and user burden are not primary concerns. Among those that did, the analysis split between qualitative mentions of issues like complexity (24.8\%, 29 papers) and formal quantification of performance costs like latency (28.2\%, 33 papers). 

This methodological split is evident in practice. Some studies, like Alshehri et al.~\cite{alshehri2023exploring}, conduct qualitative UX analysis, exploring negotiation behaviors, experience degradation and preference elicitation. Whereas others, like Xiao et al.~\cite{xiao2023micpro}, perform technical analyses by quantifying a low-latency codec's impact on STOI and latency. Engagement with this topic varied across PPMs. Algorithm- and system-based approaches show the highest rates of discussing negative UX, such as \textit{Differential Privacy} (70\%) and \textit{System Design} (61\%). Conversely, user-facing mechanisms show significantly less focus on potential drawbacks, such as \textit{Dynamic Notification} (38\%) and \textit{Privacy Label} (33\%), likely reflecting the perception that these tools cause less friction.

Similarly, the analysis of user motivation and incentives is underexplored. Nearly half of the papers (46.2\%, 54 papers) did not discuss any positive aspects. When incentives were considered, it was most often through qualitative discussions of benefits such as improved trust (41.9\%, 49 papers), with far fewer studies offering quantitative evidence (12.0\%, 14 papers). This suggests a widespread assumption that privacy is an inherent benefit, not one requiring a distinct value proposition for user adoption. \textit{Dynamic Notification} is a significant exception, showing the highest rate of engagement (88\%) and a high rate of formal quantification of its positive impacts (44\%), suggesting that benefits are mostly measured in PPMs that enable direct user interaction.

\subsubsection{Integration and Infrastructure Requirements}

\textbf{Our analysis shows that many academic proposals are not simple add-ons but require deep, systemic integration.} A \textit{mixed} integration approach is the most common (35.9\%, 42 papers), exemplified by works like Thakkar et al.~\cite{Thakkar2022ItWP}, which analyze privacy features spanning apps, devices, and cloud functions. This is supplemented by substantial reliance on the \textit{cloud} (22.2\%, 26 papers) and deep \textit{embedded OS} access (12.0\%, 14 papers), particularly in \textit{System Design} research (24 \textit{mixed}, 13 \textit{embedded OS}). For instance, Chen et al.~\cite{chen2021indistinguishability} implemented their scheduler side-channel defense directly within the real-time Linux kernel (\textit{embedded OS}). In sharp contrast, the simple \textit{companion app} model (18.9\%, 22 papers) is a typical choice for \textit{content-based} methods like \textit{Access Control} (8 papers).

\textbf{Further complicating deployment, many PPMs necessitate additional, non-standard hardware.} \textit{Edge servers} (mentioned in 27 papers) are a common requirement, particularly for \textit{System Design} (14 papers). Similarly, \textit{extra gateways} or \textit{hubs} are frequently included in both \textit{System Design} (11 papers) and \textit{Authorization} (7 papers). For example, Chi et al.~\cite{chi2023detecting} presented IoTMediator, which explicitly requires a ``hub-based architecture'' running as a ``local mediator'' (e.g., on a Raspberry Pi) to intercept and translate all device communication. Even decentralized proposals may have dependencies. Lian et al.~\cite{lian2022decentralized} proposed a federated learning mechanism that, while peer-to-peer, still requires a server to act as a coordinator. Conversely, some designs, like the RedFlash scheme~\cite{chen2022duplicates}, are notable for explicitly avoiding external infrastructure. 

\subsubsection{Interoperability and Scalability}

Interoperability and scalability are critical, non-negotiable factors for any real-world \textit{system-based} PPMs, yet they are also the most neglected. Discussion of interoperability is minimal. Only 9.4\% (11 papers) address potential conflicts with standards like Matter or Zigbee, while the vast majority (90.6\%, 106 papers) do not. 

When addressed, the topic is discussed in several distinct ways. Some papers, such as Windl et al.~\cite{windl2023investigating}, identify it as a requirement for standardization, arguing that for tangible privacy to be effective, it must be a ``standardized solution'' ensuring ``each privacy profile fits with each smart home privacy dashboard''. Other works identify interoperability challenges as practical limitations. Do et al.~\cite{do2023powering}, for example, state their prototype requires a specific ``backscatter infrastructure'' which is ``uncommon at present ... in everyday environments'', thereby hindering practical deployment. This topic appears most often in \textit{System Design} research, yet still in only 7/69 papers, suggesting that integration with existing ecosystems is often a secondary concern.

Scalability is also largely unaddressed, with the majority of papers (88.0\%, 103 papers) omitting the topic entirely. This omission is evident even in the most prolific category, \textit{System Design}, where 53/69 papers omit the topic. Formal, quantitative measurement of scalability is exceptionally rare. It appears in only a handful of papers on \textit{System Design} (2 papers), \textit{Blockchain} (1 paper) and \textit{Differential Privacy} (1 paper). This widespread neglect leaves critical questions about the viability and performance of the proposed solutions in large-scale, real-world deployments.

\begin{leftbar}
\noindent \textbf{Key takeaways (RQ1):}

$\bullet$ Research lacks real-world evaluations.

$\bullet$ Research exhibits biases in dimensions like data lifecycle, omitting data deletion.

$\bullet$ Research neglects the consideration of deployment-related dimensions (e.g., cost, interoperability and legal compliance).
\end{leftbar}

\section{Analysis of Commercially Documented Privacy Protections}

Our analysis reveals that documented protections are unevenly distributed across the 9 PPM categories. Notably, protections related to \textit{Blockchain}, \textit{Differential Privacy}, and \textit{Trading} (all alorithm-based PPMs) are absent in any documentation. We first map the observed protections to our 9 PPM categories and granularly describe their documented features (Sec~\ref{sec:analysis}). We then analyze the threat models, data lifecycle stages, and privacy properties they address (Sec~\ref{sec:dimensional}). We finally investigate how extrinsic factors, such as product category, influence this documentation (Sec~\ref{sec:category}). 

\subsection{Protections Across PPM Categories}\label{sec:analysis}

\subsubsection{Dynamic Notification}

A subset of products (38.4\%, 33/86) documents privacy-centric dynamic notifications, which are almost exclusively limited to on-device hardware status indicators. These indicators provide immediate, physical visual cues to signal active sensor data collection. This set includes 12 smart speakers/hubs, such as the \textit{Google Nest Hub Max} which employs a ``flashing dots'' when a wake word is detected and audio transmission begins. It also includes 21 cameras/doorbells, such as \textit{Nest Cam} products that display a green status LED when the camera is \textit{``on and sending video footage to Google''.}

\textbf{An additional 23.3\% (20/86) of devices documented functional notifications that, while triggered by sensor activity, were explicitly framed as security or status alerts rather than privacy indicators.} These notifications were primarily app-based push notifications. For instance, 12 camera and doorbell products (e.g., \textit{August View}, \textit{iSmartSafe}, \textit{Xiaomi Cam}) send motion alerts to the user's phone. While this activity initiates data collection, the notification is framed as a physical security event (e.g., \textit{``someone is at the door''}) rather than a notice of data processing. Similarly, 8 smart hubs (e.g., \textit{Insteon Hub}) provide alerts for device state changes (e.g., \textit{``door opened''}), but their documentation explicitly notes these are not about \textit{``privacy-related activities''} or \textit{``personal information collection''}.

\textbf{An equally large subset of products (38.4\%, 33/86) provided no documentation of any real-time dynamic notification, relying on static, one-time consent at setup.} These devices require users to agree to broad data collection policies during installation but offer no subsequent, real-time feedback when sensors are actively capturing or transmitting data. This reveals a divergence from academic proposals. While academic research highlights \textit{Dynamic Notification} as a contextual mechanism, product documentations prioritizes signaling device state rather than communicating granular privacy implications.

\subsubsection{Tangible Privacy}
\textbf{A small group of products (18/86, 20.9\%) document \textit{tangible privacy} control by providing a dedicated physical mechanism to disable sensors.} The most common mechanism is a dedicated microphone mute button (14/86 products, 16.3\%), which documentation describes as electronically cutting power to the sensor, often confirmed by a physical indicator light (e.g., \textit{Amazon Echo Dot}). A small subset of devices (4/86, 4.7\%) features a unified, physical sensor switch (e.g., \textit{Google/Nest Hub Max}). This mechanism is documented as a hardware slide-switch that simultaneously disables both the microphone and camera, explicitly noting it ``cannot be remote-switched''. Finally, a third category (3/86, 3.5\%) consists of physical camera obfuscation. In these cases, the documentation describes either an automated mechanism where the lens retracts into the device's base (e.g., \textit{Xiaomi Cam}) or a manual ``privacy shield'' that rotates to occlude the lens (e.g., \textit{Philips Hue}, \textit{Netatmo Camera}).

\textbf{Conversely, the majority of products (64/86, 74.4\%) provide no documentation of hardware-level privacy controls, with some (18/86, 20.9\%) explicitly offering only software-based toggles.} Unlike their hardware counterparts, these software controls (e.g., in \textit{Google Home}, \textit{Reolink}) do not electronically disconnect the sensor, leaving verification of the ``off'' state opaque to the user. This reliance on software is distinct from other products that lack designed privacy mechanisms, relying instead on general-purpose physical actions like unplugging the device (15/86, 17.4\%) or using main power switches (19/86, 22.1\%), which are not specialized for privacy. Furthermore, in a few cases (4/86, 4.7\%), tangible controls are documented not as a standard feature but as an inverted ``push-to-talk'' model (e.g., \textit{Amazon Fire TV} where microphone is off by default and enabled upon intentional interaction) or are offered as optional accessories (e.g., \textit{LG Smart Cam}) or premium-tier features (e.g., \textit{Roku Voice}). 

\textbf{While industry's documentation of hardware switches reflects an understanding of user trust in verifiable control, it diverges from academic proposals focusing on assurance rather than intentionality.} Current products (e.g., \textit{Amazon Echo}, \textit{Google Nest Hub Max}) provide a verifiable means of disengagement--a switch to assure the user a sensor is off. This contrasts with academic research exploring alternative tangible interactions, such as ``off-by-default'' mechanisms that require continuous, physical user action to power and enable a sensor (e.g., ``push-to-power'')~\cite{do2023powering}. 

\subsubsection{Privacy Label}

\textbf{Standardized \textit{Privacy Labels} are largely absent from product documentation, a critical misalignment with academic advocation for user-friendly transparency.} A majority of products (76/86, 88.4\%) do not feature \textit{Privacy Labels}. Instead, manufacturers rely solely on traditional, full-text \textit{Privacy Policies} (e.g.,\textit{Roku TV}). This practice deviates from privacy labels championed in academic proposals. 

\textbf{The only documented disclosures are mandatory digital disclosures imposed by application ecosystems.} 10/86 products (11.6\%) have disclosures provided by companion apps on application stores, such as the Apple Store or Google Play. Beyond this compliance, most documented artifacts consist only of technical specifications or sensor guides that detail data collection. 

\subsubsection{Access Control}

\textbf{Commercial documentation widely features access controls, but the focus is on reactive data management and data collection modification, a distinction from the academic emphasis on preventative control~\cite{ravidas2019access}.} Among these focuses, a primary form is post-collection data management (80.2\%, 69/86). This practice signals a focus on regulatory compliance rather than preventative control, centering on granting users the rights to access, review, and remove stored data. Specifically, 48.8\% (42/86) of products document explicit mechanisms, such as in-app or web portals, for users to view, select, and delete sensitive data like voice recordings and video footage (e.g., \textit{Amazon Echo Dot}, \textit{Google Nest Cam}). Furthermore, 19.8\% (17/86) of products feature opt-out rights for specific data uses (e.g., personalized advertising) or legal rights fulfillment, reflecting adherence to data regulations (e.g., \textit{Roku TV}, \textit{Flux Bulb}).


\textbf{Beyond data deletion, substantial products (59.3\%, 51/86) focus on data collection modification, emphasizing real-time, software-based control, especially for high-sensitivity camera and TV products.} Explicit software controls allowing users to switch data-generating functions (e.g., microphone toggles) on or off are featured in 29.1\% (25/86) of devices (e.g., \textit{LG TV} opt-out of voice information). A minority of products (8.1\%, 7/86) documented rule-based features for users to visually define specific privacy zones (e.g., \textit{TP-Link}, \textit{Reolink Cam}) within a camera's field of view where data collection is forbidden. Additionally, 22.1\% (19/86) of products offer \textit{privacy modes} or single-click interaction to stop data collection, highlighting the industry trend towards explicit data control. 

\textbf{39/86 products (45.3\%) documented access and rule-based frameworks, paralleling proactive academic PPMs, to manage device access and information collection.} These documentations focus on managing \textit{who} can access the device and \textit{when} data is generated. External rule of API integration is explicitly provided in 29.1\% (25/86) of products, supporting user-defined \textit{IFTTT} rules to govern data collection (e.g., \textit{Smart Brewer}, \textit{Nest T-stat}). Furthermore, role-based access control (RBAC) features for multi-user device sharing (e.g., \textit{Admin}, \textit{Member}, \textit{Guest permissions}) are documented in 32.6\% (28/86) of devices (e.g., \textit{SwitchBot}, \textit{Xiaomi}). Finally, contextual automation rules, such as geofencing for automatically toggling a device's status, are documented in 16.3\% (14/86) of products (e.g., \textit{TP-Link}). Despite this documented presence, these controls still do not feature academic innovations.

\subsubsection{System Design}

\textbf{Commercial SHDs primarily document localized computing, encryption measures and formal design principles, marking a divergence from academic proposals that focus on novel architectures like fog computing~\cite{li2022novel}.} While research explores alternative architectures, the public documentations of some products highlights practical architectural choices that minimize cloud reliance for sensitive data, which we classify into \textit{Edge Computing}, \textit{Encryption Architecture} and \textit{Formal Design Principles}. Conversely, the majority of products (66/86 products, 76.7\%) explicitly rely on the cloud for processing, and only a minority (11/86 products, 12.8\%) document purely local processing (e.g., \textit{Reolink}, \textit{Insteon}). 

\textbf{On-device processing is the most prevalent strategy (27/86 products, 31.4\%), indicating a focus of data minimization at the source of highly sensitive sensors.} This is found primarily in audio and camera-based devices, with 85.2\% (23/27) focusing on sensitive data reduction via local hotword detection (e.g., \textit{Amazon Echo Dot}, \textit{Google Home}) and on-device AI processing for visual data (e.g., \textit{Nest Cam}). For instance, \textit{Nest Doorbell (wired, 2nd gen)} localizes detection for objects like \textit{People}, \textit{Parcel}, \textit{Animal} and \textit{Vehicle} before streaming to reduce the data exposure.

\textbf{Beyond edge processing, 17/86 (19.1\%) products documented encryption methods, indicating a move towards transparency.} Specifically, 11.6\% (10/86) of products employ data security and encryption frameworks, such as the \textit{End-to-End Encryption (E2EE)} cited by \textit{Philips Hue} and two-layer encryption (\textit{AES 128-bit} and \textit{TLS}) cited by \textit{August View}. Furthermore, 8.1\% (7/86) of devices adopt a local-first processing model. For example, the \textit{Yi Cam} employs a P2P model with the server functioning as a non-content-sharing relay, ensuring video privacy. 

\textbf{Formal privacy-by-design (PbD) methodologies are evident but rare (7.0\%, 6/86), suggesting a de-prioritization of top-down systemic framework.} This high-level initiative is evident among major firms like \textit{Apple}, \textit{GE}, \textit{LG} and \textit{Samsung}. For example, \textit{Apple} demonstrates a core PbD philosophy, utilizing on-device processing and advanced architectures like the \textit{Secure Enclave}. 

\subsubsection{Authorization}

\textbf{Commercial products primarily document mandatory cloud-mediated user account registration, as opposed to novel authorization algorithms proposed in academic research~\cite{tong2022ccap,wang2022dag}.} Nearly all analyzed products (85/86, 98.8\%) explicitly or implicitly require a username/password login as the authorization gateway for device control. This mandatory registration is often required to use ``smart'' features. Examples include the explicit requirement for an account (e.g., \textit{Appkettle} and \textit{Icsee Doorbell}), and the reliance on a \textit{Samsung} account for the \textit{SmartThings Hub}. 

\textbf{Complicated authorization mechanisms such as two-factor authentication (2FA) and biometric verification are evident but sparsely documented (16.3\%, 14/86).} This practice is concentrated in high-stakes applications across major ecosystems (\textit{Amazon}, \textit{Google}, \textit{Philips}), including devices like the \textit{Nest Cam with floodlight}, \textit{Nest Doorbell (wired)}, and \textit{Philips Hue Hub}. Products from \textit{Apple} also document the use of \textit{Secure Enclave} and \textit{Keychain}. Biometric verification is highlighted in 5/86 products (5.8\%), such as \textit{August View} (biometric verified access).

\textbf{A subset of products documents nascent granular authorization mechanisms, such as data sharing permissions and RBAC, particularly for shared home environments.} Explicit user consent for sharing sensor data or granting third-party application access is documented in 11.6\% (10/86) of products. For instance, \textit{Nest Cam} and \textit{Echo Spot} state they \textit{``will only share audio/video ... if you ... explicitly give us permission''}. Furthermore, multi-user sharing or RBAC systems are documented in 32.6\% (28/86) of products. Examples include the formal `Home' concept used by \textit{Google Nest Hub/Mini}, which assigns broad rights to \textit{``family members''} and limited access to \textit{``Guest Mode''}. This shows an emergent pattern for providing configurable, multi-user authorization capabilities, particularly aligning with academic literature~\cite{zeng2019understanding}. 

\subsection{Dimensional Analysis of Protections}\label{sec:dimensional}

\subsubsection{Threat Model}

\textbf{We find that only 15/86 (17.4\%) products provide any threat model documentation, contrasting to that 76\% papers provided at least an implicit model.} These commercial documentations do not mention structured threat models in academic research (e.g., STRIDE, LINDDUN). Instead, product documentations consist of qualitative descriptions of mechanisms found across privacy policies and whitepapers. These descriptions center on specific technical controls rather than systematization of threats. For example, manufacturers emphasize encryption as a primary control, citing \textit{``data encryption''} (\textit{Meross}), default \textit{``end-to-end encryption''} (\textit{Philips Hue}), or storage in \textit{``secure, encrypted server[s]''} (\textit{Microsoft}).

\textbf{All threat model articulations are implicit models within privacy-by-design architectures, defining threats as cloud-centric data breaches, unauthorized provider access, or re-identification risks.} The documented defenses aligned with this model, emphasizing \textit{local-first processing} to prevent cloud-exposure--such as \textit{``on-device camera sensing''} (\textit{Google Nest}) and facial recognition data \textit{``never uploaded''} (\textit{SwitchBot}). Similarly, defenses included data de-identification, with \textit{Microsoft} detailing its \textit{``no data trace''} architecture, and access controls like \textit{``two-factor authentication (2FA)''} (\textit{Philips}) to mitigate unauthorized access.

\textbf{This industrial focus also parallels academia, where these defended features map to those PPM categories with clearly defined threat models: \textit{System Design} and \textit{Authorization}}. Conversely, the threats relevant to content-based PPMs (\textit{Privacy Labels}, \textit{Dynamic Notification})--which academia itself models less formally--remain unaddressed in product documentation. This suggests that the availability of well-defined academic threat models may influence industrial defense documentation.

\subsubsection{Data Lifecycle}

\textbf{We find that the data lifecycle stages prioritized in product documentation diverge substantially from those emphasized in academic literature (Table~\ref{tab:ppm_summary}).} Product documentation most frequently addresses \textit{collection} (79/86 products), primarily describing activation triggers (e.g., voice hotwords or explicit user action). The is followed by \textit{processing} (74/86), \textit{deletion} (69/86) and \textit{storage} (65/86). While \textit{processing} is also a focus in academic work (89.7\% of papers), the emphasis diverges. Product descriptions often distinguish between cloud operations and privacy-enhancing on-device computation (e.g., \textit{Nest Doorbell} processing \textit{``what it sees''} on-device). In contrast, academic literature concentrates on systemic security, such as securing model inference with trusted enclaves~\cite{hou2021model}.

\textbf{Furthermore, two misalignments emerge in the \textit{deletion} and \textit{transmission} data lifecycle stages.} First, \textit{deletion}, the most neglected stage in academia (29.9\%, 35 papers), is highly common in product documentation (69/86 products), where manufacturers emphasize user agency via granular controls (e.g., deleting records \textit{``by date range''}) and a formal \textit{``Right to Deletion''}. Conversely, \textit{transmission}, the dominant topic in academia (92.3\%, 108 papers), receives the least explicit mention in documentation (39/86 products). When mentioned, these descriptions focus on specific security guarantees (e.g., \textit{``AES-128 and TLS encryption''}) rather than the broad systemic concerns found in research.

\subsubsection{Core Privacy Properties}

\textbf{Our analysis reveals that \textit{Control} and \textit{Confidentiality} are the most frequently documented privacy properties, which aligned with their high priority in academic research.} \textit{Control} (83/86 products) is the most prevalent, manifested as tangible mechanisms (e.g., the \textit{Amazon Echo Dot}'s physical microphone button) and software-based access controls (e.g., \textit{Fire TV} data settings), mirroring its importance in research (71.8\%). \textit{Confidentiality} (71/86 products) is similarly prioritized in academia (85.5\%), though the conceptual focus diverges: product documentation emphasizes specific mechanisms like encryption (e.g., \textit{August View} using \textit{``AES 128 bit and TLS''}) and E2EE (\textit{Philips Hue Secure}), whereas research highlights system-level architectures such as federated learning.

\textbf{\textit{Transparency} and \textit{Data Minimisation} are also documented with high frequency, reflecting an emphasis consistent with academic literature.} \textit{Transparency} (70/86 products) is highlighted with static notices (e.g., privacy policies) and dynamic indicators. This use of dynamic signals (e.g., the \textit{Harman Kardon Allure's} light ring) aligns with research emphasizing content-based PPMs (53.8\%) to enhance user agency. \textit{Data minimisation} (58/86 products), which holds similar academic emphasis, is achieved in products primarily via system design, such as local-first processing (e.g., the \textit{Google Nest Hub}'s ``on-device camera sensing''), or post-collection de-identification (e.g, \textit{GE Appliances}).

\textbf{A substantial divergence from academic priorities emerges with \textit{Accountability} (15/86 products) and \textit{Anonymity} (10/86 products), which are substantially underrepresented compared to their academic focus (40.2\% and 38.5\% respectively).} When present, product accountability often manifested as user data rights (e.g., \textit{Flux Bulb}'s ``right to erasure''), while anonymity was described via data obfuscation (e.g., \textit{Apple TV}). \textit{Integrity} (5/86 products) is scarce in both domains.

\subsection{Factors Influencing Documented Protections}\label{sec:category}

\textbf{Our analysis indicates that documented protections varied by manufacturer and product category, but less so by retail price.} This variance may be attributed to the different internal technical priorities and external regulatory pressures (e.g., GDPR). For instance, brands emphasizing specific privacy features (e.g., \textit{Reolink} and \textit{Nest} with ``local-first processing'' or ``physical shutters'') provide detailed documentation, whereas others (e.g., \textit{Sengled}, \textit{Allure}) offer coarse-grained descriptions. This divergence is evident across product categories, where a device's function and the perceived sensitivity of its data influence documented protections. High-sensitivity categories like \textit{Speakers} and \textit{Cameras} often feature tangible mechanisms like physical mute buttons or shutters. Conversely, categories like \textit{Smart TVs} (\textit{Roku}, \textit{Fire TV}) are often documented as permissive by default, are fully cloud-dependent and mandate registration. 

\textbf{This inconsistency also persists within high-sensitivity categories.} Among surveillance cameras, some models (\textit{Reolink}, \textit{Yi}) prioritize local-first processing and offline storage, while others (\textit{Amazon}, \textit{Cloudcam}) remain fully cloud-dependent. Likewise, audio devices (e.g., \textit{Amazon Echo}, \textit{Google Home}) often lack documented \textit{Tangible Privacy (TP)} mechanisms, relying instead on \textit{Dynamic Notification (DN)} (i.e., light indicators) and cloud-dependent voice processing. This inconsistency extends to a single device's function. For some smart appliances (e.g., vacuums, bulbs), core non-smart functions (e.g,. cooking) are available offline, whereas all ``smart'' features are documented as cloud-dependent and require mandatory user registration. Finally, this variation exists within a single manufacturer's products, indicating that privacy decisions are highly contextual and documented at the product level.  \textit{Samsung}, for example, documents its \textit{SmartThings v2 Hub} as supporting a local AppEngine for offline automation, yet its \textit{Appliances} (e.g., \textit{Fridge}) as relegating all ``smart'' features to the cloud. These patterns underscores the importance of future product-level analysis and regulatory scrutiny. 

\textbf{In contrast, a higher price does not reliably guarantee strong documented protections.} Our dataset spans a wide price spectrum, from sub-\$50 budget devices to premium-tier appliances exceeding \$2,000. We observe that some premium-tier products (e.g., \textit{Samsung Appliances}) primarily document standardized legal disclaimers (e.g., data subject rights under GDPR or CCPA) rather than specific, technical privacy mechanisms. Conversely, several lower-priced devices document protections in a detailed manner. For instance, \textit{SwitchBot} ($<$\$120) explicitly documents that sensitive biometric data is processed and stored only locally on the device and never uploaded. Similarly, \textit{Meross} ($<$\$50) documents supports for \textit{``offline control''}. This suggests that decisions regarding documented privacy are primarily driven by factors other than cost.

\begin{leftbar}
\noindent \textbf{Key takeaways (RQ2):}

$\bullet$ Products only adopts practical mechanisms, neglecting novel PPMs.

$\bullet$ Protections highlight lifecycle stages like deletion and properties like control, deprioritizing preventative precautions and technical robustness.

$\bullet$ Documented protections varied by manufacturers and products, but not by price.
\end{leftbar}

\section{Discussions \& Open Challenges}\label{seven}

Our analysis uncovers a multifaceted gap between PPMs proposed in literature and documented in products. This disconnect presents open challenges and implications for the research, industry and policymakers.

~\\
\textbf{For academic researchers:} 
\textbf{\textit{Shifting from novelty to deployment and real-world needs.}} While continued innovation in PPMs (e.g., blockchain- or differential privacy-based PPMs) remains vital, greater attention should be given to evaluating how PPMs perform in real-world contexts. Academic work should emphasize deployability, scalability, and usability by developing frameworks that practitioners can adopt, thereby ensuring that proposed PPMs can be realistically implemented within existing device infrastructures. Collaborating closely with industry partners can further help researchers to identify deployment barriers, align privacy solutions with commercial constraints, and validate their effectiveness in real-world contexts. 

\textbf{\textit{Verifying public disclosures vs. ground-truth implementation.}} Our work analyzes public disclosures rather than ground-truth implementation. A critical open challenge is to uncover the gap between what manufacturers claim and what they actually implement~\cite{fernandes2016security,zhou2019discovering}. Future work is needed to develop and apply methods that validate public claims against observable network traffic, side-channels, or firmware analysis, moving beyond reliance on documentation. Creating shared benchmark datasets that document mismatches between claimed and observed PPMs would allow reproducibility and meta-analyses across studies.

\textbf{\textit{Uncovering manufacturer motivations and constraints.}} It is crucial to understand why manufacturers adopt or ignore PPMs, or what factors influence their postures. While initial work has explored developer perspectives in a single region~\cite{he2025privacy}, a deeper, global investigation into the technical, economic and regulatory trade-offs practitioners face is important to understand the gap for implementing PPMs. Academic research can explore motivations and constraints, offering evidence to help policymakers design incentive and accountability frameworks that align with industry practice.

\textbf{\textit{Understanding users' comprehension and response towards PPMs and their communication.}} An important open question is to assess how users comprehend and value the diverse PPMs we systematized, particularly in response to how these mechanisms are communicated. Our framework provides a foundation for this inquiry, enabling researchers to move beyond generic privacy studies~\cite{abdi2021privacy,adbi2019more}. Future work can leverage our taxonomy to design large-scale studies that measure user comprehension towards PPMs~\cite{cummings2021need} or quantify how the presence of PPMs impact consumer adoption~\cite{emami2023consumers}. 

~\\
\textbf{For industry practitioners:}
\textbf{\textit{Utilizing our classification as a framework for internal protection audits.}} Practitioners can leverage our systematization as a resource to move beyond a purely compliance-driven approach. While current public disclosures often converge on reactive mechanisms and privacy is not yet a clear market differentiator, our framework provides a tool for internal auditing. It characterizes academic privacy properties, threat models, and preventative PPMs, enabling practitioners to benchmark their internal protections against the state-of-the-art and enhance their threat modeling for risks that may not yet be addressed.

\textbf{\textit{Moving towards demonstrable privacy to build trust.}} Opaque policies are documented to pose the long-term risk which erode consumer trust~\cite{chhetri2019eliciting,emami2019exploring,tabassum2020smart,williams2017privacy}. We advocate that industry move beyond compliance and demonstrate their privacy protections more transparently. This could include publishing standardized, machine-readable privacy manifests to describe exactly what data is collected, retained and processed (locally vs. in the cloud), and integrating rigorous privacy impact assessments as part of the feature development cycle.

~\\
\textbf{For policymakers:}
\textbf{\textit{Mandating architectural transparency over textual disclosures.}} Our findings indicate that current, unstructured textual disclosures (i.e., privacy policies) are insufficient for public transparency, as they are often generic, non-specific, and difficult to access for pre-purchase assessment. Architectural transparency could be a more potent intervention, where regulators should require simple, verifiable, binary labels, akin to a nutrition label~\cite{emami2021informative}, that expose core architectural choices (e.g., ``Data Processed Locally'' vs. ``Data Processed in Cloud''). This policy shift would realign market incentives, moving the burden of discovery from consumers to manufacturers, who would be required to verifiably disclose their architecture.



\bibliographystyle{acm}
\bibliography{sample}

\appendices

\clearpage 

\section{Ethics Considerations}

We carefully considered and addressed ethical aspects throughout this research. Although our study did not involve direct experiments with human subjects, nor did it entail any harmful or deceptive designs, we adhered to established ethical principles, including Menlo Report~\cite{bailey2012menlo} and Belmont Report~\cite{beauchamp2008belmont} for our systematic literature review and analyses of publicly available commercial products.

Our analysis of scholar research and commercial devices relied solely on manually screening publicly accessible information, such as paper content, product specifications, official privacy policies, and manufacturer disclosures. We did not engage in any form of unauthorized access, reverse engineering, or data scraping that would violate terms of service or ethical standards.

We also acknowledge the broader societal implications of this work for several stakeholders. First, we acknowledge the potential impact on the commercial stakeholders named in our paper. Our categorization carries a risk of reputational harm. We mitigate this risk by clarifying that our methodology is based exclusively on publicly available documentation. Our work is an analysis of this public-facing posture, rather than a technical audit of unstated internal practices. Our aim is to contribute positively to the disclosure on digital privacy and encourage the adoption of ethically sound design practices.

Second, for consumers, we recognize that our findings may influence consumer perceptions and adoption. By highlighting the gap between academic proposals and commercial reality, our work could foster skepticism. We mitigate the risk of fostering undue alarm or a false sense of security by providing a structured, evidence-based analysis. We think that publishing and analyzing the statuses would contribute more than the potentially harmful impact, as they would make consumers informed of the products' privacy communication landscape. Our goal is to empower consumers for informed decision-making, not to deter the adoption of technology.

Third, we considered the risk that adversaries could misuse our findings. By systematizing documented weakness--such as the reactive focus on deletion rather than preventative controls or the de-prioritization of transmission protections--our work could inadvertently provide a roadmap for attackers. We mitigate this by noting that our analysis is based on publicly accessible data rather than the discovery of new, non-public vulnerabilities. We posit that the benefit to defenders (industry, researchers, and policymakers) in highlighting these gaps significantly outweighs the risks.

Fourth, for other researchers, in our analysis of the research landscape, we risk mischaracterizing the contributions of our peers. Our aggregate finding that literature often neglects deployment barriers is a characterization of a collective gap. We mitigate the risk of unfair individual critique by applying a systematic and uniformly defined coding scheme to all 117 papers. We frame the expression as neutral rather than critiquing specific papers. We aim to guide future research directions, rather than critiquing specific prior work.

All findings are reported truthfully and with appropriate citations to ensure academic accountability and support future replication efforts. We affirm that this research complies with the ethical standards.








\section{Coding Dimensions and Criteria for PPM Analysis}\label{app:dimension}

This appendix details the comprehensive codebook established for analyzing academic research papers concerning PPMs for smart home IoT devices. The coding was performed by two researchers, achieving an inter-rater reliability of 0.85 (Krippendorff's alpha). 

\subsection{Dimensions Reflecting Academic Priorities}

\textbf{A1: Threat Model}

$\bullet$ Description: The clarity, scope, and formality with which the research paper defines its assumed threat model, including attacker goals and capabilities.

$\bullet$ Coding Guideline: Ordinal scale with optional free-text. 0 (None) denotes threat model is not stated or is indiscernible. 1 (Partial) denotes threat model is formally mentioned or implied. 2 (Full) denotes threat model is formally defined with explicit attacker capabilities and assumptions. Besides the scores, the threat model should also has a brief summary of key attacker assumptions if stated.

\textbf{A2: Empirical Evaluation}

$\bullet$ Description: The primary methodology employed in the paper to evaluate the proposed PPM. 

$\bullet$ Coding Guideline: Single-choice categorical (primary type). Categories included simulation, test-bed/prototype experiment, field study/deployment, formal proof/theoretical analysis, user study (lab/controlled), dataset-based analysis, not applicable/purely conceptual.

\textbf{A3: Data Lifecycle Coverage}

$\bullet$ Description: The stages of the data lifecycle that the PPM is designed to protect.

$\bullet$ Coding Guideline: Multi-label binary. Stages included collection, transmission, storage (device), storage (cloud/server), processing, sharing/dissemination, deletion/retention.

\textbf{A4: Privacy Property Targeted}

$\bullet$ Description: The specific privacy property or principles(s) that the PPM aims to achieve or enhance.

$\bullet$ Coding Guideline: Multi-label. Properties include confidentiality, integrity, availability, anonymity, unlinkability, plausible deniability, data minimisation, transparency, user control/intervention, accountability, obfuscation, differential privacy, others (specify).

\textbf{A5: Resource Constraints}

$\bullet$ Description: Whether the research explicitly considers and quantifies the impact of resource constraints (e.g., CPU, memory, battery/energy, network bandwidth) on the PPM's performance or feasibility, particularly for resource-constrained IoT devices. 

$\bullet$ Coding Guideline: Ordinal scale. 0 (None) denotes resource constraints are not stated or considered. 1 (Partial) denotes resource constraints are qualitatively mentioned or acknowledged as a factor. 2 (Full) denotes resource constraints are formally modeled, and their impact is quantitatively evaluated. 

\subsection{Dimensions Reflecting Industry Priorities}

These dimensions were used to assess academic literature from the industry's perspectives to gauge considerations of industry factors. 

\textbf{I1: Cost}

$\bullet$ Description: The monetary, resource, or computational cost (which could be transferred to monetary cost) associated with implementing the PPM.

$\bullet$ Coding Guideline: Ordinal scale. 0 (None) denotes that the cost is not discussed. 1 (Partial) denotes that the cost is qualitatively mentioned (e.g., ``low cost'', ``computationally expensive'') without specific quantification. 2 (Full) denotes that cost is quantitatively discussed or estimated with concrete numbers or resource usage tables, figures or studies.

\textbf{I2: Legal Compliance}

$\bullet$ Description: Explicit references made to data protection / privacy standards, laws or regulations (e.g., GDPR, CCPA, PIPL, ISO standards).

$\bullet$ Coding Guideline: Nominal list; multiple entries allowed. Record each specific standard or regulation mentioned. If none, mark as `None mentioned'.

\textbf{I3: Negative User Experience (UX)}

$\bullet$ Description: Evidence that the research or product documentation acknowledges or evaluates potential negative impacts on user experience introduced by the PPM, such as increased latency, complex onboarding steps, or UI friction. 

$\bullet$ Coding Guideline: Ordinal scale. 0 (None) denotes no mention of negative user experience aspects. 1 (Partial) denotes negative user experience aspects are discussed qualitatively. 2 (Full) denotes that negative user experience aspects are formally measured or empirically evaluated (e.g., usability study metrics, performance benchmarks related to UX).

\textbf{I4: Incentive/Positive UX}

$\bullet$ Description: Whether the research or product material suggests that the PPM could act as an incentive for users, potentially improving user experience or increasing purchase intention.

$\bullet$ Coding Guideline: Ordinal scale. 0 (None) denotes no mention of incentives. 1 (Partial) denotes mentioning improvement to user experience or other non-monetary benefits as an incentive (nominal description). 2 (Full) explicitly links the PPM to increased user purchasing intention or adoption.

\textbf{I5: Integration Depth}

$\bullet$ Description: The architectural layer where the PPM or its control primarily resides in the smart home ecosystem.

$\bullet$ Coding Guideline: Single-choice categorical. Categories included firmware, embedded OS, companion app, cloud platform, gateway device, mixed/multi-layer or not specified.

\textbf{I6: Infrastructure Requirement}

$\bullet$ Description: Requirement for additional, dedicated hardware or significant infrastructure modifications beyond standard user devices or existing home networks (e.g., a dedicated gateway, edge server, or a federated learning coordinator). 

$\bullet$ Coding Guideline: Binary with optional free-text. Yes or No. If Yes, specify the type of extra infrastructure required in free-text. 

\textbf{I7: Interoperability Conflicts}

$\bullet$ Description: Any documented or discussed conflicts, limitations or challenges related to the PPM's interoperability with existing smart home standards, protocols, or ecosystems (e.g., Zigbee, Z-Wave, Matter, Thread, Alexa, Google Home).

$\bullet$ Coding Guideline: Binary with optional free-text. Yes or No. If Yes, describe the nature of the conflict in free-text.

\textbf{I8: Scalability Evidence}

$\bullet$ Description: The extent to which the research provides evidence (e.g., through experiments or simulations) of the PPM's ability to scale, for instance, with an increasing number of devices, users, or data volume (e.g., evaluation involving $>$100 devices or $>$1 million messages).

$\bullet$ Coding Guideline: Ordinal scale. 0 (None) denotes scalability is not mentioned or evaluated. 1 (Partial) denotes scalability is mentioned or discussed qualitatively. 2 (Full) denotes scalability is formally measured or demonstrated through empirical evaluation with defined metrics. 

\begin{table}[!htbp]
\centering
\caption{List of Smart Home Devices (SHDs) Analyzed in This Study.}
\label{tbl:implementation_situation}
\resizebox{0.462\textwidth}{!}{%
\begin{tabular}{|l|l|p{4.5cm}|} 
\hline
\textbf{Manufacturer} & \textbf{Model} & \textbf{Products} \\
\hline
Allure & Audio & Allure with Alexa \\ \hline
\multirow{3}{*}{Amazon} & Camera & Cloudcam \\
 & Audio & Echo Dot, Echo Spot, Echo Plus \\
 & TV & Fire TV \\ \hline
Flux & Automation & Flux Blub \\ \hline
Anova & Appliances & Anova Sousvide \\ \hline
Apple & TV & Apple TV \\ \hline
Appkettle & Appliances & Appkettle \\ \hline
August & Camera & August View \\ \hline
GE & Appliances & GE Microwave \\ \hline
\multirow{5}{*}{Google} & Hub & Nest Hub \\
 & Camera & Nest Cam, Nest Cam IQ, Nest Cam (battery), Nest Cam with floodlight, Nest Cam (indoor, wired), Nest Doorbell (wired)/Nest Hello, Nest Doorbell (wired, 2nd gen), Nest Doorbell (battery), and Nest Hub Max \\
 & Speaker & Google Home, Google Home Mini, Google Home Max, Google Nest Hub, Google Nest Hub (2nd gen), Google Nest Hub Max, Google Nest Mini, Google Nest Audio \\
 & Surveillance & Nest Cam, Nest Cam IQ, Nest Cam (battery), Nest Cam with floodlight, Nest Cam (indoor, wired), Nest Doorbell (wired)/Nest Hello, Nest Doorbell (wired, 2nd gen), and Google Nest Doorbell (battery) \\
 & Chromecast & Chromecast with Google TV (HD) and Chromecast Voice Remote \\ \hline
Honeywall & Automation & Honeywall T-Stat \\ \hline
Icsee & Camera & Icsee Doorbell \\ \hline
Insteon & Hub & Insteon Hub \\ \hline
iSmartSafe & Camera & iSmartSafe \\ \hline
LifeLock & Camera & LifeLock \\ \hline
Lightify & Hub & Lightify Hub \\ \hline
LG & TV & LG TV \\ \hline
Magichome & Automation & Magichome Strip \\ \hline
Meross & Automation & Meross Door Opener \\ \hline
Microsoft & Audio & Invoke \\ \hline
\multirow{3}{*}{Nest} & Automation & Nest T-stat \\
 & Camera & Nest cam \\
 & Camera & Nest cam outdoor \\ \hline
Netatmo & Appliances & Netatmo Weather \\ \hline
\multirow{2}{*}{Philips Hue} & Hub & Philips Hue Hub \\ 
 & Automation & Philips Bulb \\ \hline
Reolink & Camera & Reolink Cam \\ \hline
Roku & TV & Roku TV \\ \hline
\multirow{3}{*}{Samsung} & Appliances & Samsung Dryer \\
 & Appliances & Samsung Fridge \\ 
 & TV & Samsung TV \\ \hline
Sengled & Hub & Sengled Hub \\ \hline
\multirow{3}{*}{Smarter} & Appliances & Smarter Brewer \\ 
 & Appliances & Coffee Machine \\
 & Appliances & iKettle \\ \hline
SmartThings & Hub & SmartThings Hub \\ \hline
\multirow{2}{*}{Smartlife} & Automation & Smartlife Bulb \\
 & Automation & Smartlife Remote \\ \hline
\multirow{2}{*}{SwitchBot} & Hub & SwitchBot \\ 
 & Appliances & SwitchBotCurtain, SwitchBotBlindTilt, SwitchBotMiniRobotVaccum \\ \hline
TP-Link & Automation & TP-Link Bulb, TP-Link Plug \\ \hline
Uniden & Camera & Ubell Doorbell \\ \hline
WeMo & Automation & WeMo Plug \\ \hline
Wink 2 & Hub & Wink 2 Hub \\ \hline
\multirow{5}{*}{Xiaomi} & Camera & Xiaomi Cam \\
 & Hub & Xiaomi Hub \\
 & Automation & Xiaomi Strip \\
 & Plug & Xiaomi Plug \\
 & Appliances & Rice Cooker \\ \hline
Yi & Camera & Yi Cam \\ \hline
Zmodo & Camera & Zmodo Doorbell \\ \hline
\end{tabular}
}%
\end{table}

\section{Coding Dimensions for Public Stated PPMs}\label{app:product_dimension}

To analyze the 86 commercial products, we coded dimensions in two categories. The first, \textbf{Product and Market Context}, captures product identity and economic positioning. The second, \textit{Academic-Oriented Dimensions}, was adapted from our literature review dimensions (Appendix~\ref{app:dimension}) to facilitate a direct comparison between academic proposals and commercial disclosures. We selected \textit{A1: Threat Model}, \textit{A3: Data Lifecycle Coverage}, and \textit{A4: Privacy Property Targeted} as they are central to a PPM's focus and are often described in documentation. We excluded \textit{A2: Empirical Evaluation} and \textit{A5: Resource Constraints} because public disclosures do not typically provide this level of technical detail. Similarly, industry-oriented dimensions (e.g., \textit{I1: Cost}, \textit{I8: Scalability}) were excluded, as our analysis is limited to public documentation, which generally provides insufficient detail to verify claims regarding these factors. Besides, industry-oriented dimensions are typically naturally fulfilled within products.

\textbf{\textit{Product and Market Context}} We analyzed dimensions related to the product's identity, economic positioning and regional markets, which collectively inform the commercial and regulatory background.

$\bullet$ \textit{P1: Price.} Recorded each product's listed price as an indicator of its market position, a relevant metric given that prior work confirms consumers' willingness to pay a premium for enhanced privacy~\cite{emami2023consumers}.

$\bullet$ \textit{P2: Primary market.} Identified each product's geographical primary market to contextualize its design choices in light of regional regulations and cultural expectations, an analytical dimension recognized as important in prior research~\cite{kumar2019all}. 

$\bullet$ \textit{P3: Brand's country of origin.} Recorded the parent company's country of origin, as this factor indicates the governing legal framework and can influence user perceptions of privacy and security~\cite{protick2024unveiling,vetrivel2023examining}. 

\textbf{\textit{Academic-Oriented Evaluation Dimensions}} We chose \textit{A1: Threat Model}, \textit{A3: Data Lifecycle Coverage} and \textit{A4: Privacy Property Targeted}, with the coding criteria similar to those used for evaluating academic papers. 

$\bullet$ \textbf{P4: Threat model.} This dimension adopted the original A1, except that we only differentiated those mentioning the threat model (coded as 1) from those not mentioning (coded as 0), because there were hardly any quantification of the threat model in products.

$\bullet$ \textbf{P5: Data lifecycle coverage.} This dimension adopted the original A3 and used the same classification criteria. A stage was coded as covered only if the product documentation explicitly described a protection mechanism targeting that stage.

$\bullet$ \textbf{P6: Privacy property targeted.} This adopted the original A4 with the same classification criteria. Similarly, a property was coded as covered only if the documentation explicitly mentioned a protection that reflected that property.

\end{document}